\documentclass[11pt, oneside]{article}   	
\usepackage{geometry}                		
\usepackage[parfill]{parskip}    		
\usepackage{graphicx}				
\usepackage{amssymb}
\usepackage{natbib}
\usepackage{rotating}
\usepackage{url}

\title{Stylised Choropleth Maps for New Zealand Regions and District Health Boards}
\author{Thomas Lumley}

\begin{document}
\maketitle
\begin{abstract}
New Zealand has two top-level sets of administrative divisions: the District Health Boards and the Regions. In this note I describe a hexagonal layout for creating stylised maps of these divisions, and using colour, size, and triangular subdivisions to compare data between divisions and across multiple variables. I present an implementation in the {\sf DHBins} package for R using both base graphics and {\sf ggplot2}; the concepts and specific hexagonal layout could be used in any software. 
\end{abstract}

\section{Introduction}

Maps are a popular way to display data that come with a geographical/spatial reference.  Choropleth maps, in which regions are shaded according to the data value, are simple but have well-known weaknesses, especially for data summarising the human population.  Areas with higher population density tend to have spatially smaller administrative units; they are visually less salient, and may even be hard to see.  Cartograms, which scale the area to show the data value, overcome this problem at the cost of reducing the recognisability of the map.  In this note we present a design for an intermediate case where the map is distorted to make all the area units easily visible and coloured as a choropleth map.  

Important prior examples of this idea include the `state visibility map' of \citet{monmonier-map} and the `statebins' from  \emph{The Washington Post} \citep{statebins}. 
Monmonier's map of the US expands the eastern states and compresses the western states, but retains all the boundaries between states and the general orientation of the boundaries.  

The `statebins' map replaces each US state by a square, taking advantage of the roughly rectangular layout of many US states.  In contrast to the state visibility map, it does not preserve all the state boundaries, especially in the northeast, but together with state labels it does provide enough geographic information to identify individual states easily.  Stylised choropleth maps such as these are especially useful as small multiples \citep[p67]{tufte}, where the effort of geographic lookup is needed just once for a set of maps: an example is the micromaps of \citet{micromaps}.  

A related class of prior examples is hexagonal maps used to show electorates. These present many of the same design issues, but have the additional feature (and constraint) that all the area units have the same population, making them closer to cartograms in some ways.

In this note I present stylised choropleth layouts for New Zealand top-level divisions: the District Health Boards and Regions, using a hexagonal rather than a rectangular structure. First, I will describe the Regions and District Health Boards and the types of data where these comparisons may be helpful. In section~\ref{DHBins-code} I present an implementation of the maps in R, using both base graphics and {\sf ggplot2} \citep{R-itself,ggplot2}.  In section~\ref{triangles} I describe how multi-class data can be displayed on the stylised choropleth map, by dividing each hex into triangles.

\subsection{Regions and District Health Boards}

New Zealand, unlike Australia, Canada, and the United States, is not a federation; the subnational levels of government have only the powers they are delegated from the national government.  The top-level subnational government units are Regional Councils. There are nine Regions in the North Island and seven in the South Island. Some of these contain Territorial Authorities and District or City Councils, others (such as Auckland) are Unified Authorities combining multiple levels.
Local government expenditure at all levels was about NZ\$9 billion in 2015-16.  The lower-level units of government are not necessarily nested in the higher-level ones; for example, Taup\=o District is split between four Regional Councils, along watershed divisions.

Health funding is allocated to a different set of units, the District Health Boards. In 2018--19 this funding came to NZ\$18.3 billion. There are 15 DHBs in the North Island and five in the South Island.
Some of these are closely aligned with regions --- Northland Region and Northland DHB, or Gisborne Region and Tair\={a}whiti DHB ---
 but the DHBs have somewhat less variation in population than the Regions.  

\section{Region and DHB choropleths in R}
\label{DHBins-code}

The {\sf DHBins} R package is available from \url{github.com/tslumley/DHBins}. It provides hexagonal layouts for the District Health Board areas and Regions. The functions  \verb"regionbin()" and \verb"dhbin()" draw the hexagonal layouts using base R graphics, and the functions \verb"geom_region()" and \verb"geom_dhb" provide the corresponding {\sf ggplot2} geoms.  The functions take an argument giving the radius (relative to the full hexagons) and the colour of each hex.

\begin{sidewaysfigure}[p]
\includegraphics[height=0.75\textheight]{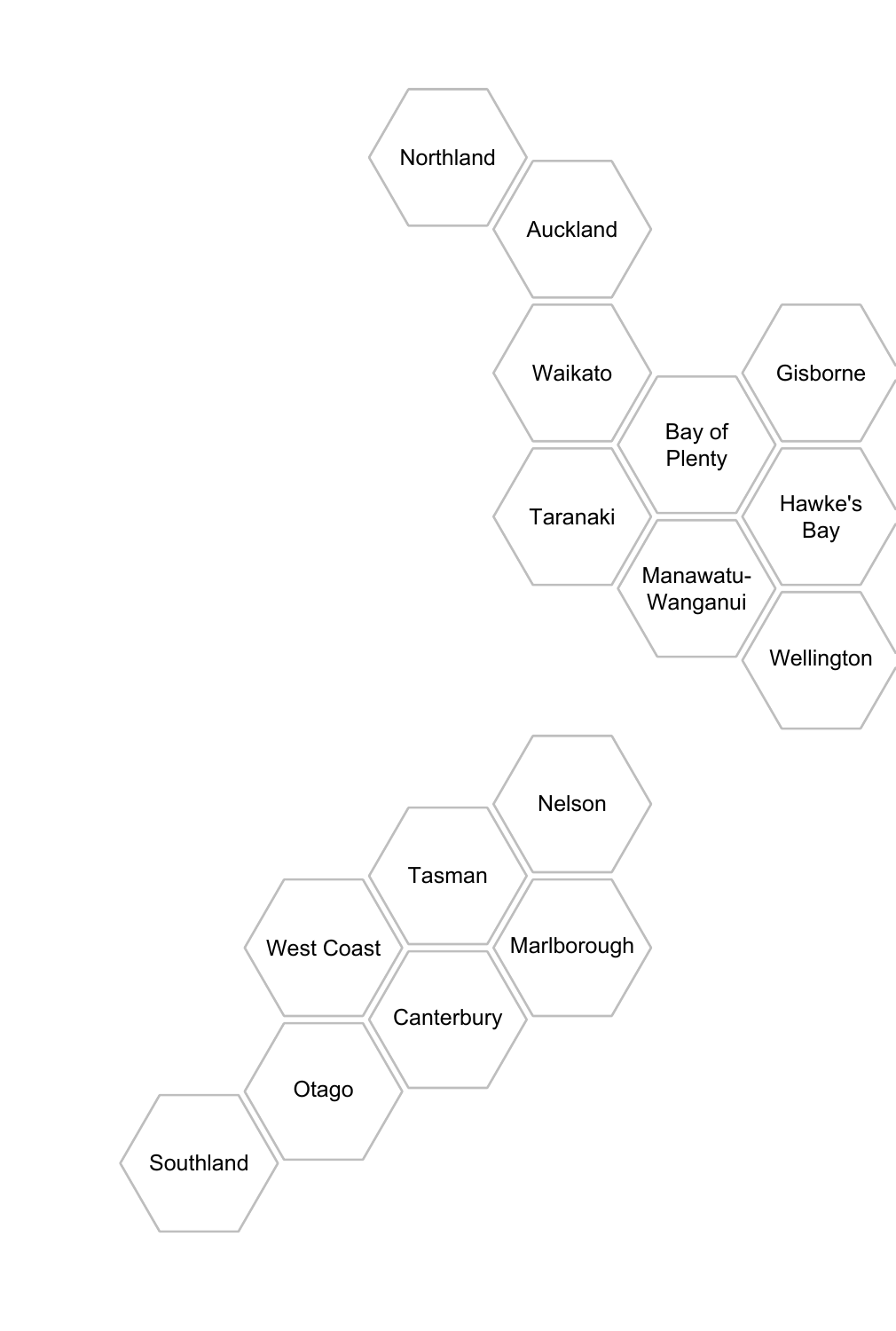}
\includegraphics[height=0.75\textheight]{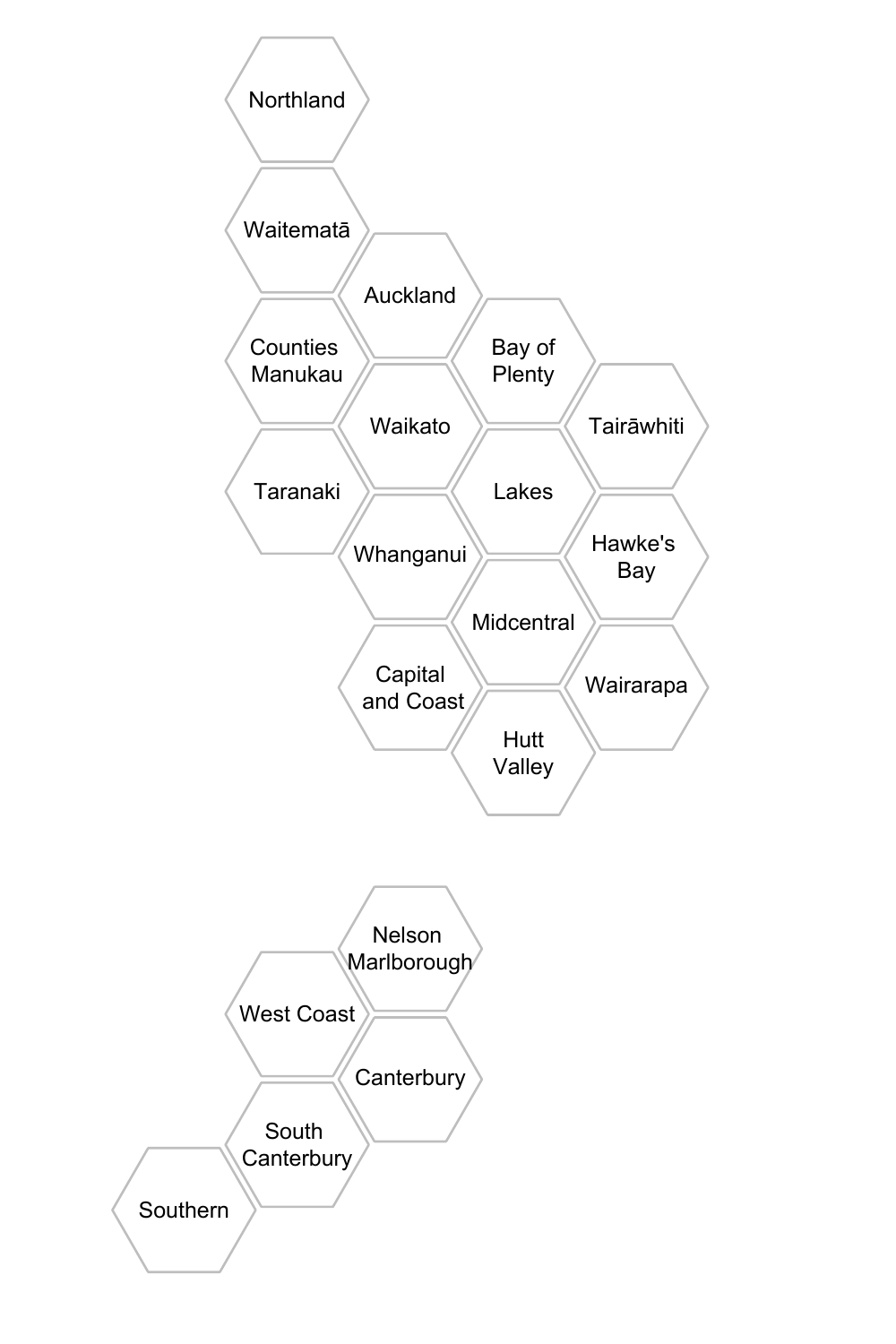}

    \caption{Regions (left) and District Health Boards (right), created with base graphics}
    \label{big}
\end{sidewaysfigure}

Figure~\ref{big} shows the two hexagonal layouts, with the unit names, drawn with base graphics.   Both layouts capture the relative orientation of the areas, and most of the intersections with the coast. However, the MidCentral District Health Board is incorrectly shown as landlocked, and the West Coast region/DHB are not shown as contiguous with the Southern DHB and Southland Region.  The layouts also capture the extreme points of the geography reasonably well, although the DHB map rounds off East Cape and the region map shows the north coast of the South Island as convex at Nelson rather than concave.

\begin{figure}[ht]
\includegraphics[width=0.48\textwidth]{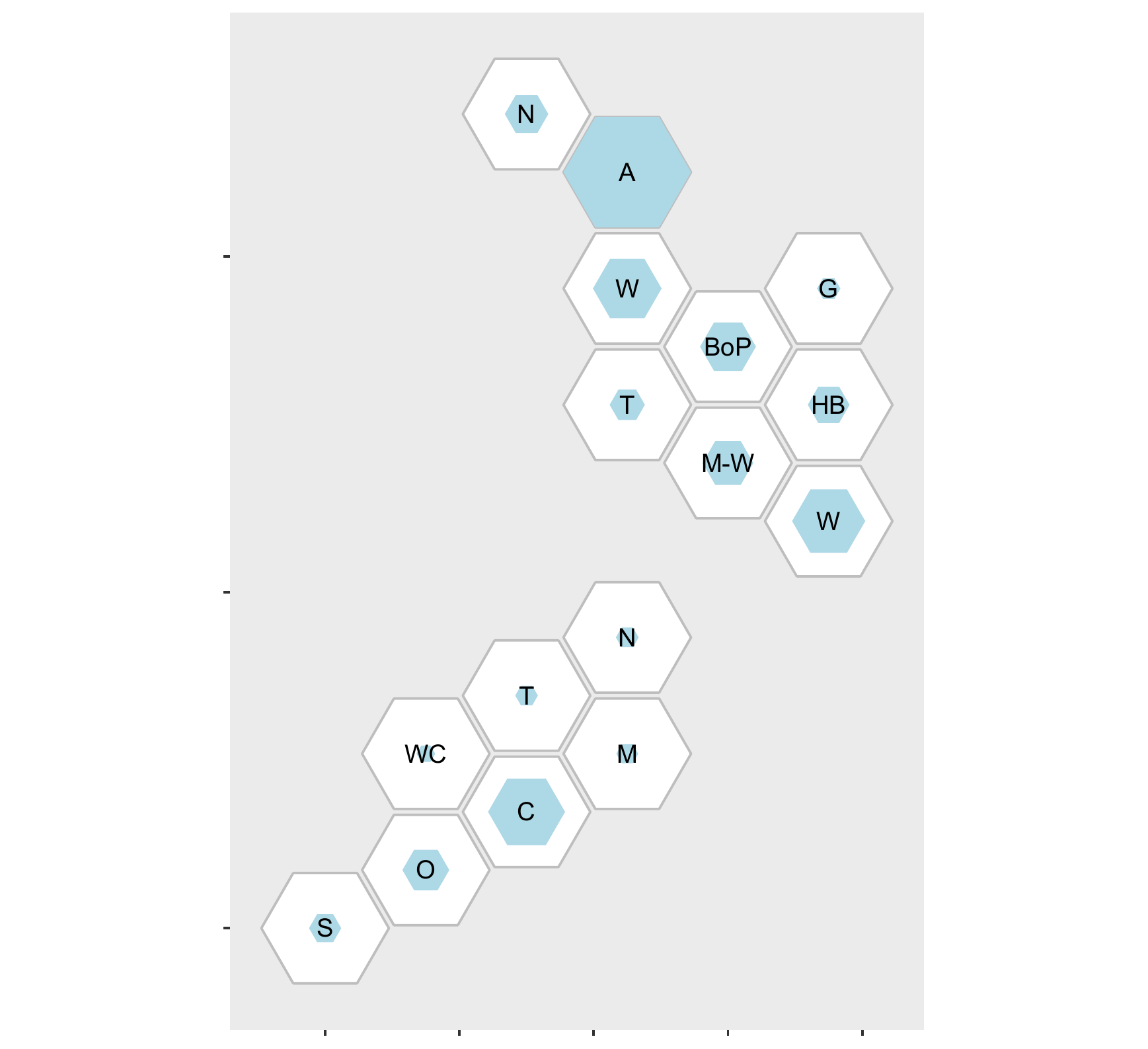}
\includegraphics[width=0.48\textwidth]{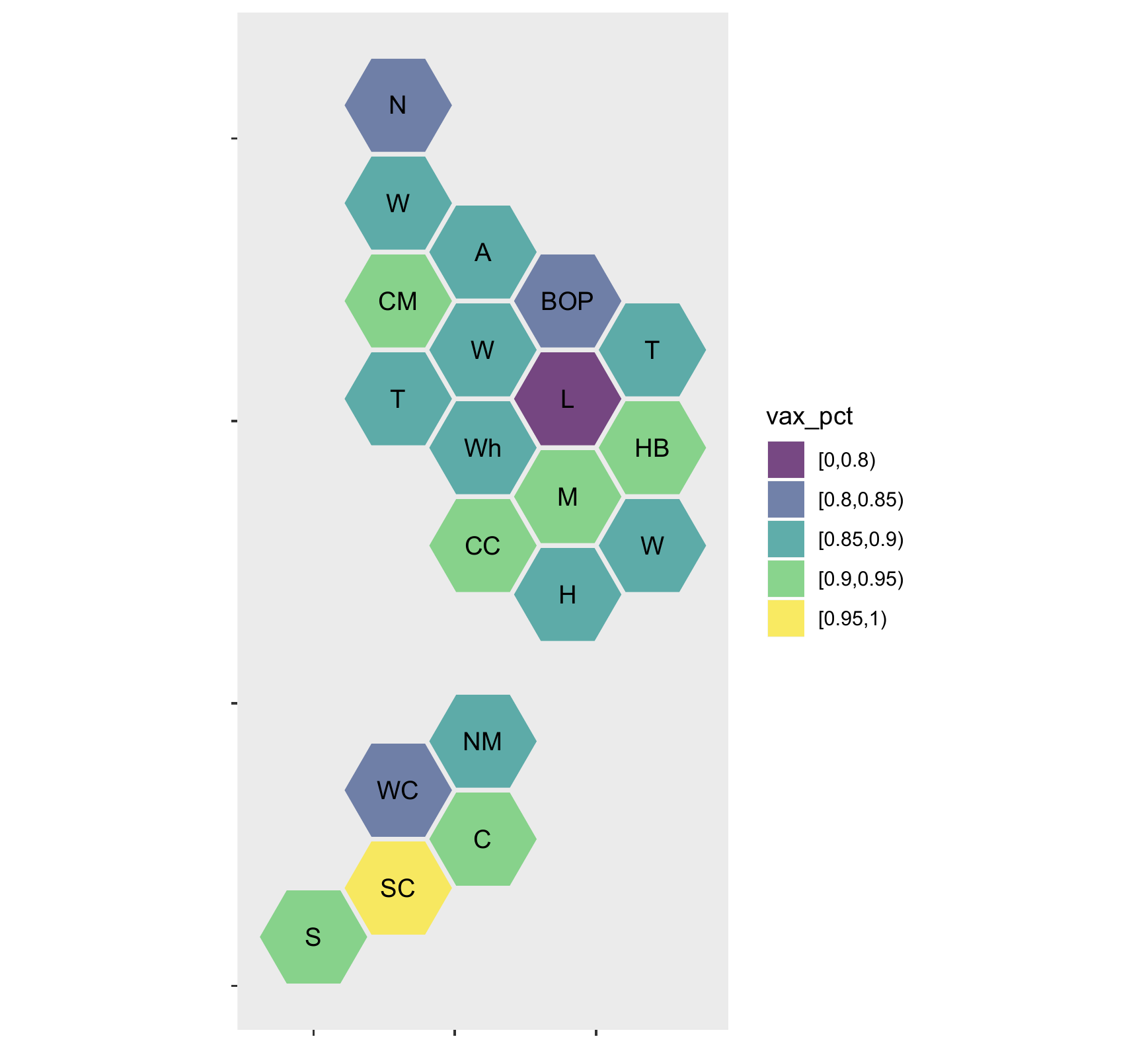}

    \caption{Regions (left) and District Health Boards (right), with abbreviated names, using {\sf ggplot2}. The region areas are proportional to the population of the area. The DHB areas are shaded according the proportion of 5-year-old children having all their recommended immunisations.}
    \label{small}
\end{figure}

Figure~\ref{small} shows the two layouts with abbreviated names. The abbreviated names are not unique, but in likely applications of the maps it would be implausible for users to confuse, eg, Waikato and Wairarapa DHBs or Taranaki and Tasman Regions. In fact, it may often be sufficient to have one map with names and leave unlabelled any subsequent maps in the same document.

 In the region map, the area of each hex is proportional to the population of the region (so the radius is proportional to the square root of population).  In the DHB map, the shading depends on the proportion of 5-year-old children who had completed all their appropriate vaccinations, for the third quarter of 2019 \citep{immune}.  The highest category corresponds to the official quality target (and to the herd-immunity threshold needed for measles). There is one DHB in this category, South Canterbury, and even they are right on the boundary.   As with all the data sets used in this note, the data are included in the {\sf DHBins} package.

Some care is needed with colour selection if labels are to be used; typically, we would would want a colour scale with a wide range of luminance, but labelling with white or black requires a narrower range of luminance. 

\subsection{Facets}
\begin{figure}[ht]
\includegraphics[width=\textwidth]{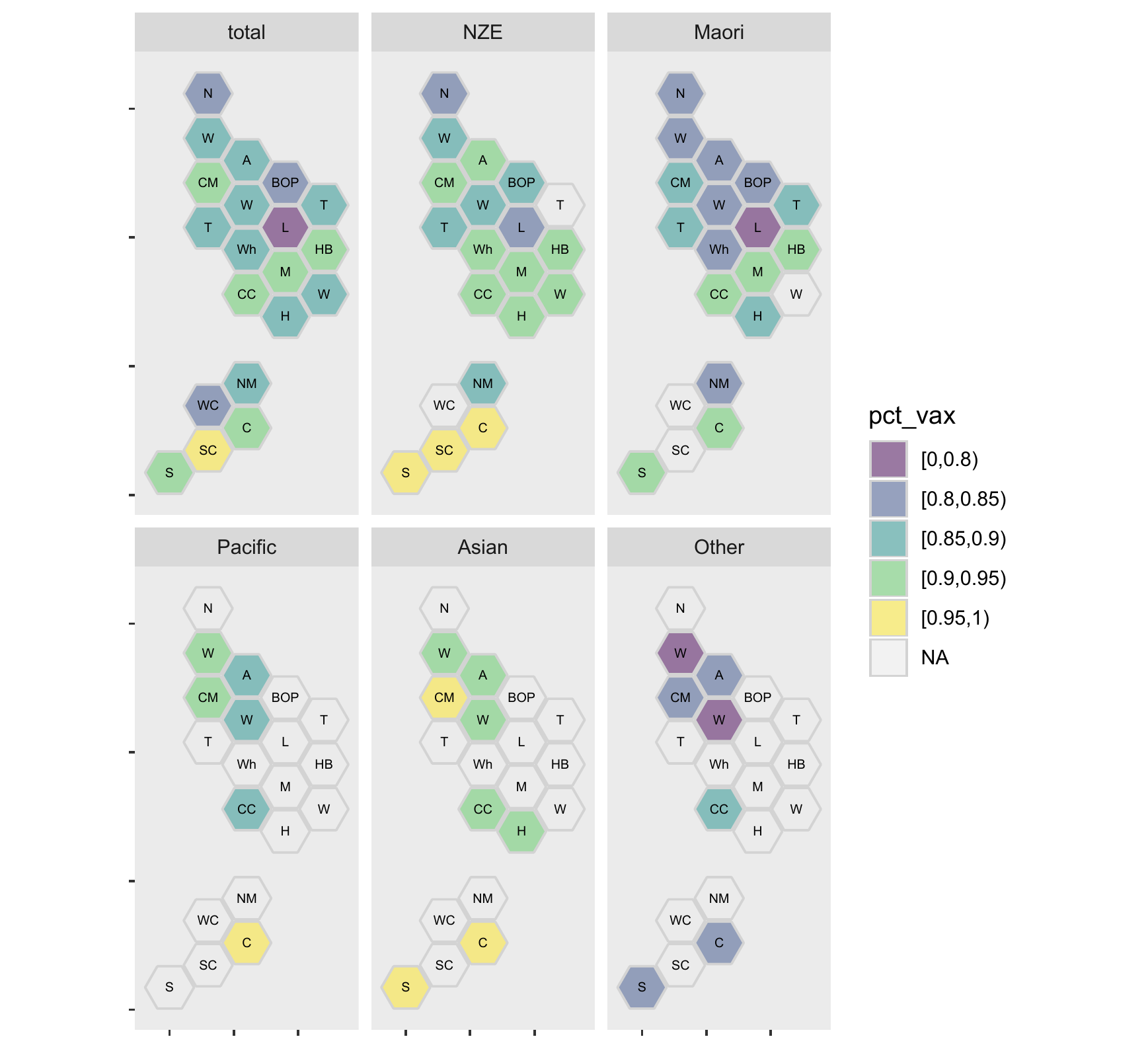}

    \caption{The DHB areas are shaded according the proportion of 5-year-old children having all their recommended immunisations, and are faceted by top-level ethnicity categorisation. Empty regions have insufficient data to estimate the proportions. `NZE' is `European, including New Zealand European'}
    \label{facet}
\end{figure}

Stylised choropleth maps are of most use for making 'small multiples' comparisons across variables, where the distortions in the map do not confound the comparisons.  Figure~\ref{facet} shows immunisation coverage by District Health Board, split up by ethnicity. Regions with data on fewer than 70 children from an ethnic category are set to missing. `NZE' is `European, including New Zealand European'. 

 It is clear there is some variation in DHBs --- for example, Lakes DHB, and to a lesser extent Northland DHB has low vaccination coverage in all groups, whereas Canterbury and Southland have higher coverage. There is also ethnic variation: all DHBs do worse at providing vaccination to M\=aori children than Asian and NZ European children. 

\section{Triangular subdivision}
\label{triangles}
Size and shading allow a single value to be displayed for each area unit. In order to display categorical data, we divide each hex into six triangles and aim to assign $[6p]$ triangles to a category with proportion $p$ of the total.  Simply rounding $6p$ to the nearest integer does not guarantee that the number of triangles assigned will add up to six; we use the Webster/Saint-Lagu\"e algorithm to allocate exactly six triangles.  Triangular subdivision has previously been suggested for hexagonally-binned scatterplots \citep{hextri-package}.

These triangle maps are superficially related to pie charts, but they do not share the deficiencies of pie charts.  The standard objection to pie charts is the difficulty of judging angles; in these triangle maps the angles are discrete and the number of possibilities is small, so counting triangles is fast and reliable. Conversely, an important limitation of the triangular subdivisions is their discreteness: they cannot display a category with probability 1/12th or less. More generally, they will be biased according to whether the category proportion is rounded up or down. 

In the R implementation, I provide a function \verb"tri_alloc()" for allocating triangles based on category counts, base-R functions \verb"dhbtri()" and \verb"regiontri()", and {\sf ggplot2} functions \verb"geom_dhb_tri()" and \verb"geom_region_tri()". The base-R functions work on `wide' format data, with a six-column matrix giving the triangle assignments.  The {\sf ggplot2} geoms work on `long' format data, with a single column of colours and an identifier to map them to triangles.

\begin{figure}[ht]
\begin{center}
\includegraphics[height=0.4\textheight]{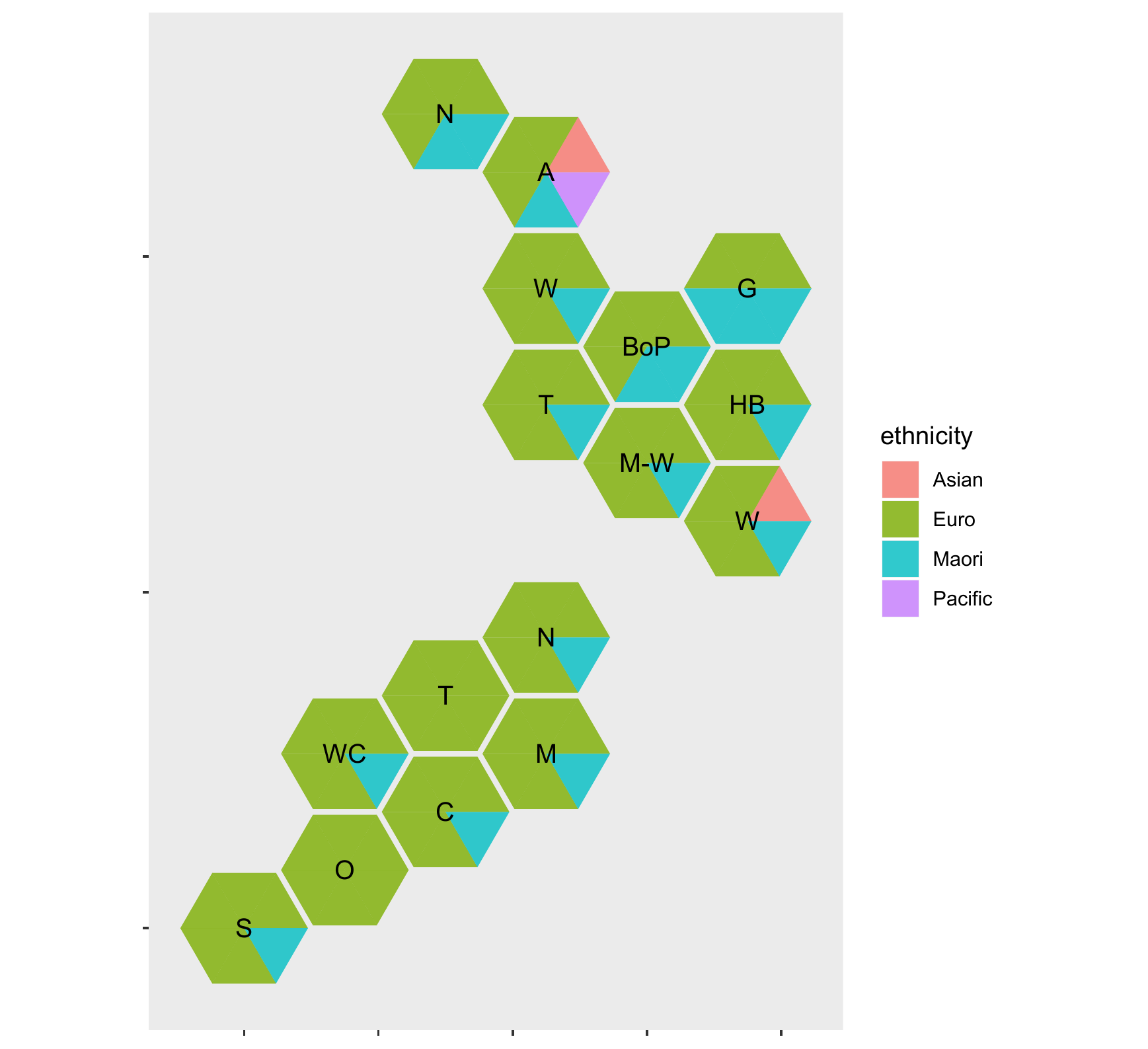}
\end{center}
    \caption{Distribution of ethnicity by Region. `Euro' includes European and NZ European. Numbers for Middle Eastern, Latin American, and African, and Other ethnicities were too small to appear. }
    \label{hextri}
\end{figure}

Figure~\ref{hextri} shows the distribution of four ethnic groups across the Regions.  Auckland and Wellington have substantial Asian and Pacific populations. The M\=aori population varies from zero-sixths  in Otago and Tasman to three-sixths in Gisborne.  The 50\% figure for M\=aori in Gisborne is an example of the impact of rounding: the actual figure is 42\%, which is very close to 2.5/6.

\section{Discussion}

I have proposed an approach to choropleth maps for New Zealand administrative units, facilitating small-multiple comparisons of variables across Regional Councils or District Health Boards, with equal visual weight given to each unit. Triangular partitioning of hexagonal maps also allows limited amounts of multi-category data to be displayed. I have also provided an implementation of the maps in R.

This approach can be applied to other countries, but the optimal tessellation and its usefulness will depend on the physical and social geography of the country.  For example, in the related context of hexagonal cartograms,  the website \url{fivethirtyeight.com} has used hexagonal maps for US and UK electorates, but obtaining a hexagonal map for Australian electorates (\url{150hexagons.com}) involves giving up on between-state boundaries.

\bibliographystyle{apalike} 
\bibliography{dhbins.bib}

\end{document}